\documentclass[11pt,twoside]{article}

\usepackage{asp2021}

\usepackage{xcolor}

\aspSuppressVolSlug
\resetcounters

\begin{document}

\title{PulSKASim: A Pulsar Simulator for SKA-Scale Interferometric Observations}

\author{Xiaotong~Li$^1$ and Vladislav~Stolyarov$^2$}
\affil{$^1$University of Oxford, Oxford, United Kingdom; \email{egbdfmusic1@gmail.com}}
\affil{$^2$University of Cambridge, Cambridge, United Kingdom; \email{vs237@cam.ac.uk}}

\paperauthor{Xiaotong~Li}{egbdfmusic1@gmail.com}{0009-0006-4959-0015}{University of Oxford}{Department of Engineering Science}{Oxford}{Oxfordshire}{OX1 3QG}{United Kingdom}
\paperauthor{Vladislav~Stolyarov}{vs237@cam.ac.uk}{0000-0001-8151-828X}{University of Cambridge}{Cavendish laboratory}{Cambridge}{Cambridgeshire}{CB3 0HE}{United Kingdom}



\begin{abstract}
Accurate simulation of pulsar flux variability is critical for testing Square Kilometre Array (SKA) interferometric pipelines. However, most existing simulators neglect the effects of integration time and related observational parameters, limiting their realism and utility for interferometric end-to-end testing. To address these shortcomings, we develop a Pulsar Simulator for SKA-scale interferometric observations (PulSKASim), which models pulsar flux evolution across pulsar period, maximum flux, duty cycle, and noise, accounting for integration time, sampling, and observation duration, and naturally models flux smoothing that arises from finite integration within each dump time. PulSKASim generates synthetic measurement sets using functions in simulators for radio interferometers, such as OSKAR and Pyuvsim, where each snapshot contains pulsars with controlled flux levels, enabling realistic per-time-slot experiments. This simulator allows for detailed assessment of calibration, imaging, and detection pipelines under realistic SKA-like conditions, bridging pulsar variability modelling with interferometric simulation in a way not achievable by existing tools.
\end{abstract}



\section{Introduction}

Pulsars are rotating, highly magnetised neutron stars whose beams produce pulsing signals to observers on Earth. They are valuable probes of strong-field gravity, ultra-dense matter, and the interstellar medium, and form a core component of the Square Kilometre Array (SKA) science programme\footnote{\url{https://www.skao.int/en/explore/science-goals}}. Accurate pulsar simulations are essential for designing observing strategies, validating signal-processing pipelines, and preparing for next-generation radio telescopes, such as the SKA. Existing pulsar simulators, such as PsrSimSig\footnote{\url{https://psrsigsim.readthedocs.io/en/latest/}} and Radio Pulsar Signal Generator \citep{rpsg}, are not designed for producing interferometric Measurement Sets (MSs) with pulsar variability. 

To fill this gap, we develop a simulator, called PulSKASim, which supports SKA-scale configurations and enables testing of interferometric pulsar search, imaging, and timing pipelines under realistic conditions. PulSKASim is open source and is available on GitHub \footnote{\url{https://github.com/egbdfX/PulSKASim/}}.

\section{Methodology}

The structure of PulSKASim is shown in Fig. \ref{diagram}. It consists of two major components, a flux generator and a radio interferometric (RI) simulator. The flux generator produces a flux sequence with one value per snapshot, given the pulsar period $T$, maximum pulsar amplitude $A$, sampling period (telescope dump time) $T_s$, pulsar duty cycle $D$, total simulated duration $T_\mathrm{sim}$, and noise standard deviation $\sigma_n$. This sequence is inputted into an RI simulator, such as OSKAR\footnote{\url{https://ska-telescope.gitlab.io/sim/oskar/}}~\citep{oskarieee} or Pyuvsim\footnote{\url{https://pyuvsim.readthedocs.io/en/latest/index.html}}~\citep{pyuvsim}. Simulated MSs are generated by specifying the observing frequency, phase centre, pulsar position, observation time, dump time, and telescope layout to the RI simulator.
\articlefigure[width=0.9\textwidth]{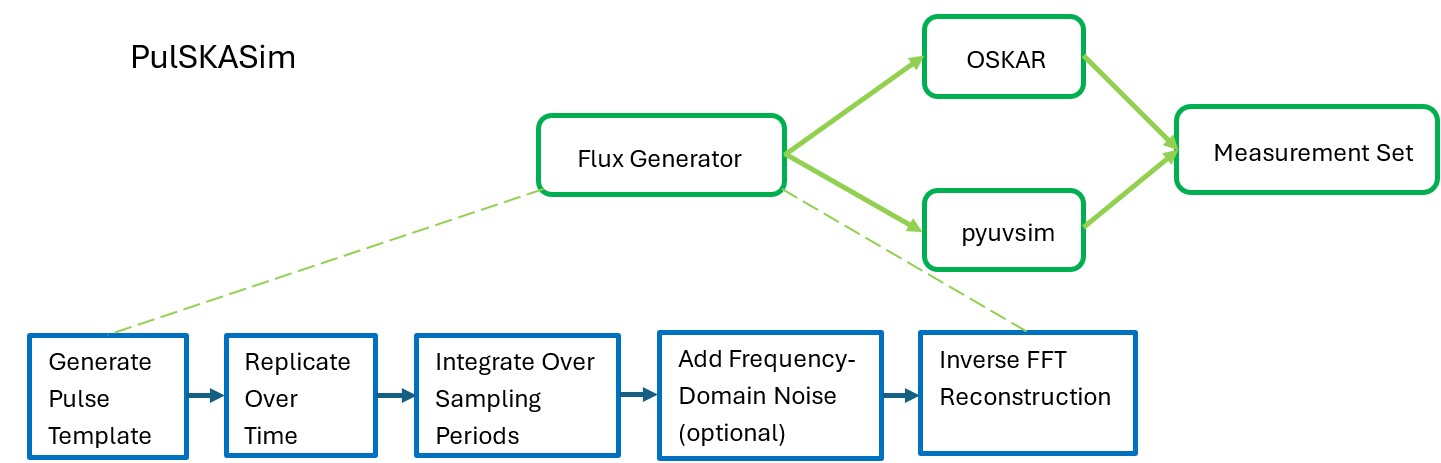}{diagram}{Structure of PulSKASim.}

Specifically, the flux generator first creates a pulsar template, modelling the pulsar profile as a Gaussian \citep{Pulsar} with width determined by $D$. The pulsar template is offset-corrected so that it smoothly connects to zero outside the duty cycle and is then replicated over the total simulated duration to produce a continuous periodic signal. To mimic interferometer integration, instantaneous brightness is integrated over each sampling period to produce time-averaged flux values. Optionally, to simulate realistic noise, random noise is introduced into the Fourier domain, controlled by $\sigma_n$. The flux is then reconstructed in the time domain via inverse Fourier transform, and the resulting sequential flux values are stored in the output catalogue.

The flux generator subsequently triggers the RI simulator. In this work, we consider OSKAR, a GPU-accelerated RI simulator optimised for large arrays such as the SKA. For time-variable flux realisations, each snapshot is simulated as an individual MS. Each simulation can be run in parallel across multiple allocated nodes. The resulting snapshot-level datasets can then be merged into a single MS using \textit{casacore} functions. The merged MS is the output of PulSKASim. Alternatively, we employ Pyuvsim, a lightweight, Python-based simulator designed for accurate visibility calculations and algorithm testing in radio astronomy. Similar to OSKAR, to simulate time-varying sources by Pyuvsim, users must run separate simulations for each time step with updated sky models; these runs can be parallelised with MPI, and the resulting datasets (usually in \textit{uvh5} format) should be merged afterwards using \textit{pyuvdata} or \textit{casacore} and converted to the MS.

\section{Performance}
\subsection{Fidelity}
The fidelity of PulSKASim depends largely on the performance of the flux generator. To assess it, we compare the simulated pulsar flux with the observation of PSR J0901-4046 \citep{realpul2} as an example. A total of 1500 snapshots were generated in this experiment. Figure \ref{fidelity} (Left) shows the frequency spectrum of the real pulsar alongside the noise-free simulated spectrum. In it, the real signal exhibits a prominent low-frequency component, which corresponds to low-frequency noise. This is evident in Fig. \ref{fidelity} (Middle), where removing the low-frequency component reveals the high-pass filtered signal. Furthermore, as shown in Fig. \ref{fidelity} (Right), by adding the same noise pattern as the real signal, the frequency spectrum of the simulated signal closely matches that of the real signal, demonstrating the high fidelity of the simulation.

\articlefigure[width=0.8\textwidth]{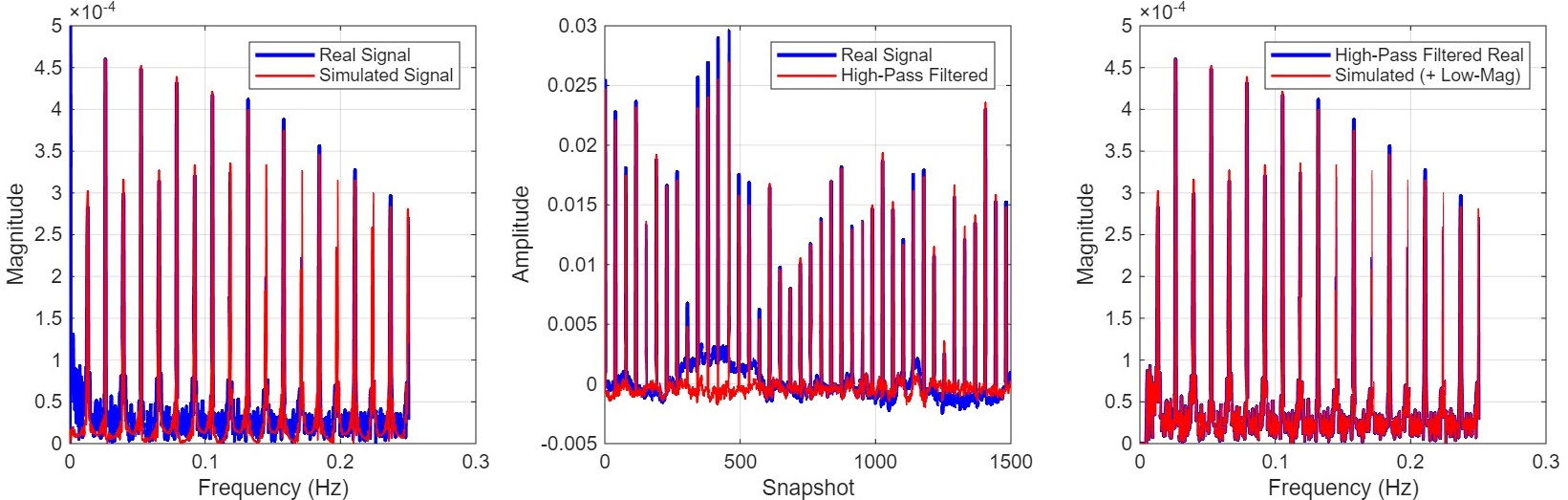}{fidelity}{Fidelity assessment. \emph{Left:}  Frequency spectra of PSR J0901-4046 and the noise-free PulSKASim simulation ($T=75.886\mathrm{s}$, $T_s=1.999\mathrm{s}$, $D=1.4/360$, $T_\mathrm{sim}=3000\mathrm{s}$, $\sigma_n=0.0$). \emph{Middle:} Comparison of the original and high-pass filtered signals, showing the effect of low-frequency removal. \emph{Right:} Frequency spectrum of the simulated signal with real-noise pattern compared to that of the real signal.}

\subsection{Robustness}
The robustness of PulSKASim is evaluated under different Nyquist sampling conditions by varying the ratio between $T_s$ and $T$. The results indicate that the simulator behaves reliably across oversampling, critical sampling, and undersampling cases. PulSKASim is a radio interferometric simulator designed for pulsar observations. It can be used to generate synthetic data for testing Fast Imaging Pipelines (FIPs) \citep{fipska,fiptoi}. Such image-domain detection pipelines are particularly effective for identifying long-period transients, which are challenging to detect using traditional time-domain approaches. In this case, the transients are typically oversampled. Furthermore, these FIPs can also be applied to short-period transients: although undersampling introduces frequency aliasing, the transients can still be successfully localised in the image domain. Consequently, the robustness of PulSKASim ensures that it is applicable to testing both oversampled and undersampled science cases.

\subsection{Computational Efficiency}
To evaluate the computational efficiency of PulSKASim, we perform simulations using the parameters of PSR J2251-3711 \citep{p2251} and the MeerKAT telescope model as an example. The median time per snapshot for simulations with different number of frequency channels is shown in Fig. \ref{timing}. Since Pyuvsim is designed as reference software that prioritises high accuracy, it is not expected to deliver high computational performance in simulations. It relies on MPI for parallelisation, and its execution time depends significantly on the number of CPU cores used. We tested both simulators on a consumer desktop equipped with a 6-core AMD Ryzen 5600X CPU and an NVIDIA RTX 3060 GPU. Our tests show that, to match the performance achieved by OSKAR using a GPU with approximately 3000 CUDA cores, Pyuvsim must be run on an HPC compute node with 100-200 CPU cores.

\articlefigure[width=0.7\textwidth]{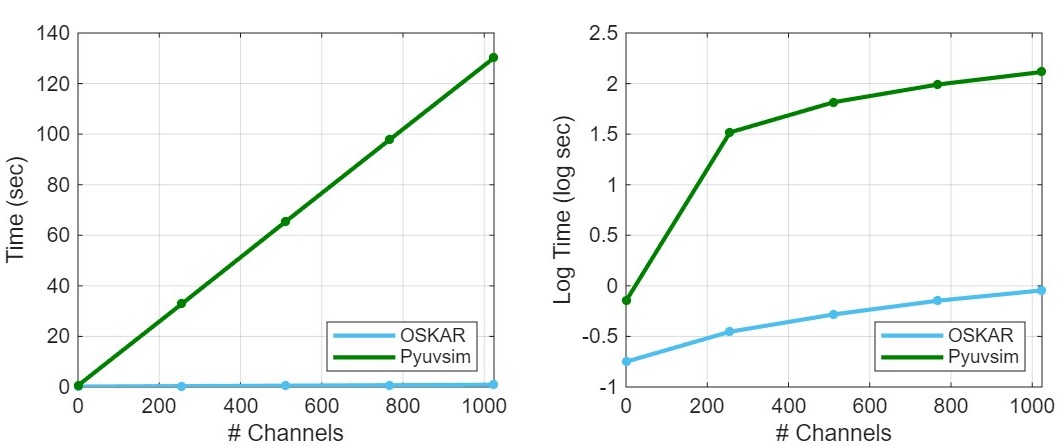}{timing}{Computation times in (\emph{Left}) linear and (\emph{Right}) logarithmic representations.}

\section{Conclusions}
We propose PulSKASim, an open-source simulator capable of producing realistic, time-variable pulsar signals in interferometric measurements. Our results show high fidelity to real pulsar observations, robustness across a range of sampling conditions, and scalable computational performance. PulSKASim fills a critical gap in pulsar research, providing a tool for designing observing strategies, validating pipelines, and preparing for next-generation radio telescopes.

\acknowledgements The authors would like to acknowledge the use of the ATNF Pulsar Catalogue\footnote{\url{https://www.atnf.csiro.au/research/pulsar/psrcat}} \citep{atnf}.

\bibliography{adass2025.bib}  

@book{Pulsar,
  title = "{Handbook of Pulsar Astronomy}",
  author = {D. Lorimer and M. Kramer},
  year = {2004},
  month = {December},
  publisher = {Cambridge University Press},
  series = {Cambridge Observing Handbooks for Research Astronomers}
}

@inproceedings{fipska,
  author = "Stolyarov, Vladislav  and  Li, Xiaotong  and  Heywood, Ian  and  Adamek, Karel",
  title = "{A fast imaging pipeline for transient detection in interferometric data}",
  doi = "10.22323/1.514.0009",
  booktitle = "PoS(HEASA2025)",
  year = 2025,
  volume = "514",
  pages = "009"
}

@INPROCEEDINGS{oskarieee,
  author={Mort, Benjamin J. and Dulwich, Fred and Salvini, Stefano and Adami, Kristian Zarb and Jones, Mike E.},
  booktitle={IEEE PAST}, 
  title={OSKAR: Simulating digital beamforming for the SKA aperture array}, 
  year={2010},
  volume={},
  number={},
  pages={690-694},
  doi={10.1109/ARRAY.2010.5613289}}

@INPROCEEDINGS{rpsg,
  author={Santos, João and Brito, Diogo and Tavares, Gonçalo and Fernandes, Jorge},
  booktitle={2018 IEEE ISCAS}, 
  title={Radio Pulsar Signal Generator}, 
  year={2018},
  volume={},
  number={},
  pages={1-4},
  doi={10.1109/ISCAS.2018.8351229}}

@article{atnf,
year = {2005},
month = {apr},
publisher = {},
volume = {129},
number = {4},
pages = {1993},
author = {Manchester, R. N. and Hobbs, G. B. and Teoh, A. and Hobbs, M.},
title = {The Australia Telescope National Facility Pulsar Catalogue},
journal = {AJ},
}

@article{p2251,
    author = {Morello, V and Keane, E F and Enoto, T and others},
    title = {The SUrvey for Pulsars and Extragalactic Radio Bursts - IV. Discovery and polarimetry of a 12.1-s radio pulsar},
    journal = {MNRAS},
    volume = {493},
    number = {1},
    pages = {1165-1177},
    year = {2020},
    month = {02},
    }

@article{realpul2,
	Author = {Caleb, M. and Heywood, I. and Rajwade, K. and others},
	Journal = {Nat. Astron.},
	Number = {},
	Pages = {828-836},
	Title = "{Discovery of a radio-emitting neutron star with an ultra-long spin period of 76 s}",
	Volume = {6},
	Year = {2022}
}

@misc{fiptoi,
      title="{FIP-TOI: Fast imaging pipeline for pulsar localisation with a transient-oriented radio astronomical imager}", 
      author={X. Li and K. Adamek and M. Giles and W. Armour},
      year={2025},
      eprint={2512.06254},
      archivePrefix={arXiv},
      primaryClass={astro-ph.IM}, 
}

@article{pyuvsim, 
    year = {2019}, 
    publisher = {The Open Journal}, 
    volume = {4}, 
    number = {37}, 
    pages = {1234}, 
    author = {Lanman, Adam E. and Hazelton, Bryna J. and Jacobs, Daniel C. and Kolopanis, Matthew J. and Pober, Jonathan C. and Aguirre, James E. and Thyagarajan, Nithyanandan}, 
    title = {pyuvsim: A comprehensive simulation package for radio interferometers in python}, journal = {J. Open Source Softw.} 
}


\end{document}